\documentclass[twocolumn,prl,aps]{revtex4-1}
\usepackage{graphicx}
\usepackage{epstopdf}

\newcommand{\e}{{\rm e}}

\newcommand{\ld}{{\lambda}}

\newcommand{\bea}{\begin{eqnarray}}
\newcommand{\eea}{\end{eqnarray}}
\newcommand{\be}{\begin{equation}}
\newcommand{\ee}{\end{equation}}
\newcommand{\ba}{\begin{eqnarray}}
\newcommand{\ea}{\end{eqnarray}}

\newcommand{\nn}{\nonumber}
\newcommand{\la}{\label}
\newcommand{\w}{\Omega}
\begin{document}
\title{Maximal intensity higher-order Akhmediev breathers of the nonlinear Schr\"odinger equation and their systematic generation}

\author{Siu A. Chin$^1$, Omar A. Ashour$^{1,2}$, Stanko N. Nikoli\'c$^{2,3}$, Milivoj R. Beli\'c$^2$ }
\affiliation{$^1$Department of Physics and Astronomy,
Texas A\&M University, College Station, TX 77843, USA}
\affiliation{$^2$Science Program, Texas A\&M University at Qatar,
P.O. Box 23874 Doha, Qatar}
\affiliation{$^3$Institute of Physics, University of Belgrade,
Pregrevica 118, 11080 Belgrade, Serbia}

\begin{abstract}

\noindent 
It is well known that Akhmediev breathers of the nonlinear cubic 
Schr\"odinger equation can be superposed nonlinearly via the
Darboux transformation to yield breathers of higher order. Surprisingly, we find that
the peak height of each Akhmediev breather only adds {\it linearly} to form
the peak height of the final breather. Using this new peak-height formula, 
we show that at any given periodicity, there exist a unique high-order breather 
of maximal intensity. Moreover, these high-order breathers form a continuous hierarchy, 
growing in intensity with increasing periodicity. For any such higher-order breather, 
a simple initial wave function can be extracted from the Darboux transformation
to dynamically generate that breather from the nonlinear Schr\"odinger equation.

\end{abstract}


\maketitle

The study of high-intensity optical solitions on a finite background, known as
``breathers" (and ``rogue waves"), is of growing importance in modern nonlinear 
photonics. For a comprehensive reference of recent works, see the review by 
Dudley {\it et al.} \cite{dud14}. One way of achieving high intensity is to create
higher-order versions of these breathers. (We regard rogue waves as special 
cases of breathers with infinite period \cite{akh09}.) While it has been know for 
a long time \cite{akh88} that these higher-order breathers can be composed 
from first-order breathers via the 
Darboux transformation (DT), the recursive complexity of the 
transformation \cite{akh88,akh093,ank11,ked112,ked12} has obscured insights 
into the working of DT's nonlinear superposition. In this work, we 
find analytically that, despite the nonlinear superposition, the 
peak heights of the breathers only add {\it linearly}. From this key result,
one can prove that: I) At each periodicity, there is a unique
higher order breather of {\it maximal} peak intensity. II) With increasing
periodicity, these higher-order breathers form a continuous hierarchy 
of single-peak solitary waves with monotonically rising intensity. 
III) In the limit of an infinite period, these higher-order breathers
morph smoothly into rational rogue waves of the same order. 
IV) From the breather's wave function generated by DT, 
a simple initial wave function can be extracted to {\it dynamically} regenerate 
that high-order breather in the nonlinear Schr\"odinger equation (NLSE).
Since the NLSE is an excellent model for propagating light pulses in an optical fiber, 
our results strongly suggest that breathers of extreme intensity and short duration 
can be systematically produced in optical fibers. 
 
Let's first summarize some well-known properties of first-order Akhmediev breathers (ABs) \cite{akh86,akh87}.
The breather's wave function
\ba
&&\psi(t,x)=\biggl[\!1\!+\! \frac{2(1\!-\!2a)\cosh(\ld t)\!+\!i\ld \sinh(\ld t)}
{\sqrt{2a}\cos(\w x)-\cosh(\ld t)}\!\biggr]\!\e^{it},\ \ \
\la{ab}
\ea
is an exact solution to the cubic NLSE
\be
i\frac{\partial\psi}{\partial t}
+\frac12 \frac{\partial^2\psi}{\partial x^2}+|\psi|^2\psi=0,
\la{sch}
\ee
on a finite background: $|\psi(t\!\rightarrow\!\pm\infty,x)|\rightarrow 1$.
Its most fundamental characteristic is that it is
{\it periodic} over a length $L$ parametrized by the 
modulation parameter $a$:
$$
L={\pi}/{\sqrt{1-2a}}.
$$
Only in the singular limit of $a\!\rightarrow\!1/2$, $L\!\rightarrow\!\infty$,
that it becomes the non-periodic, Peregrine soliton \cite{per83}. 

At a given $a$, because of this basic periodicity, the allowed Fourier modes can 
only have wave numbers
 \be
 k_m=m\Omega \qquad{\rm for}\quad m=0,\pm1,\pm2, \cdots .
 \ee
where $\Omega$ is the interval's fundamental wave number:
\be
 \Omega=2\pi/L=2\sqrt{1-2a}.
\la{omg}
 \ee
The growth factor 
$
\ld=\sqrt{8a(1-2a)}=\w\sqrt{1-(\w/2)^2}
$
is due to the instability of this fundamental mode, as determined by the Bogoliubov 
spectrum \cite{bog47,fet71} or by the Benjamin-Feir \cite{ben67} instability. 
This growth factor, when real, implies that all modes with $|k_m|<2$ are unstable. 
Specifically, the first nonzero $|m|$ harmonic modes are unstable for $|m|\w<2$, 
or at the parameter values \cite{erk11}
\be
a>a^*_m\equiv\frac12 (1-\frac1{m^2}).
\ee
If $a$ {\it were} negative, then (\ref{omg}) implies that $\w>2$ and all
modes are stable. A negative $a$ therefore corresponds to a stable, plane Stokes wave.

The AB wave function (\ref{ab}) peaks at $t=0$, with profile 
\be
\psi(0,x)=1+\frac{2(1-2a)}{\sqrt{2a}\cos(\Omega x)-1}.
\la{abk}
\ee
The maximum peak height is at $x=0$,   
\be
|\psi|_{max}=1+2\sqrt{2a}.
\la{abkh}
\ee
As $a$ increases from $0$ to $1/2$, this peak height increases from the background height 
of 1 and smoothly matches the Peregrine's \cite{per83} peak height of 3. 

By using DT, an $n$th-order breather can be constructed from $n$ ABs with an arbitrary
set of real modulation parameters 
\be
a_1>a_2>a_3>\cdots a_n>0.
\la{na}
\ee
(All $t$- and $x$-shift parameters are set to zero.)   
However, such a construction would overlook the importance of periodicity. 
Given an initial AB with $a_1=a$, it is periodic over a length of $L_1=L$.
For any $a_{k}<a_1$, the resulting $\w_k$, 
if incommensurate with $\w$, would completely destroy the periodicity of the original AB. 
Even if $\w_k$ were commensurate with $\w$, unless $\w_k$ is just a
multiple of $\w$, the periodic length $L$ must be enlarged to accommodate both wave numbers.
While there is no logical argument forbidding such an arbitrary $a_k$ construction, 
it is reasonable to insist that the higher-order breather retains the same periodic 
length $L$ as the initial AB. 
In this case, one must choose
$\w_k=k\w$, resulting in the following set of modulation parameters: 
\be 
a_k=k^2(a-\frac12)+\frac12.
\la{mak}
\ee

\begin{figure}
		\includegraphics[width=0.95\linewidth]{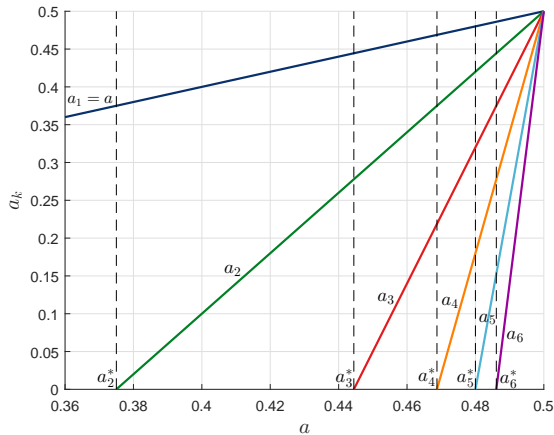}
       \caption{Modulation parameters $a_k$ of (\ref{mak}) as functions of $a$. 
Vertical broken lines indicate the locations of $a_m^*$.
       } 
\la{aks}
\end{figure}

This set of $a_k$ as a function of $a$ is plotted in Fig. \ref{aks}. 
Note that $a_k>0$ only when $a>a_k^*$. Let's denote the region $a_n^*<\!a<\!a_{n+1}^*$ as $R_n$; 
then in each region $R_n$ there are exactly $n$ values of (\ref{na}) 
that can be combined by DT to form an $n$th-order breather. We will show below that
such a breather has the highest peak intensity at any given periodic length parametrized by $a$. 

The Darboux transformation gives
\be
\psi_n(t,x)=\psi_{n-1}(t,x)+\delta\psi(t,x),
\ee
where $\delta\psi(t,x)$ depends {\it recursively} on all the previous-order wave functions \cite{akh88,akh093,ank11,ked112}.
This is classic nonlinear superposition. However, we will prove in the Appendix that
for an $n$th-order AB,
the maximum peak height at $t\!=\!0$ and $x\!=\!0$ only adds {\it linearly}, given by
\be
|\psi|_{max}=1+2\sum_{k=1}^n\sqrt{2 a_k}.
\la{htmax}
\ee
This is the key, new finding of this work, valid for an arbitrary set of real $a_k$. For
the set of $a_k$ given by (\ref{mak}), we can further deduce that:
1) In each region of $R_n$ there is a unique $n$th-order AB with peak height given by (\ref{htmax}).
This peak height is maximal because it is a sum over all available and possible $a_k$'s of a given periodic length. 
2) At regions lower than $R_n$, this $n$th-order AB does not exist because some required $a_k$ are not positive. 
3) At regions greater than $R_n$, this $n$th-order AB retains the highest peak height among all
$n$th-order ABs. For example, in $R_3$, we have $a_1>a_2>a_3>0$. 
Clearly, from Fig. \ref{aks} and (\ref{htmax}), the second-order AB formed 
from $a_1$ and $a_2$ will have the greater peak height than the AB2 formed 
from $a_1$ and $a_3$ or $a_2$ and $a_3$. The last case also illustrates that
the peak height of an AB2 formed from any two $a$ values having commensurate
wave numbers will always be lower than that formed from 
wave numbers $k_1$ and $k_2$ over the {\it same} periodic interval. 
Therefore, by (\ref{htmax}), the $n$th-order AB formed from the first $n$ values of (\ref{mak}) 
has peak intensity greater than any other AB having the same periodic length.

\begin{figure}
		\includegraphics[width=0.95\linewidth]{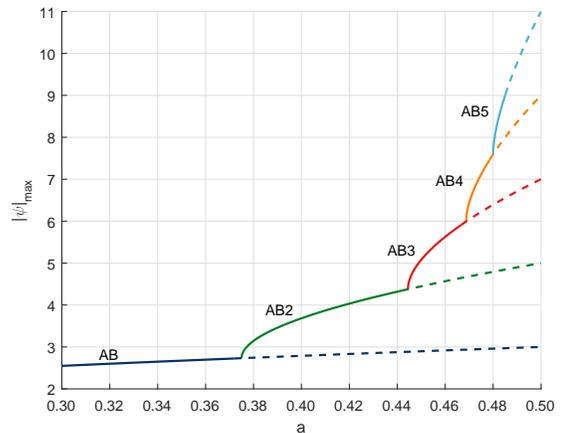}
       \caption{Peak heights of maximal intensity $n$th-order Akhmediev breathers
at any spatial periodic length parametrized by the modulation parameter $a$.
} 
\la{rwh}
\end{figure}

The peak heights of these maximal higher-order ABs are plotted in Fig. \ref{rwh}. 
At each $R_n$ region, the maximal intensity breather is indicated as a solid
line. These solid lines can be joined continously over each region,
forming a single hierarchy of maximal intensity breathers.
At higher $R_n$ regions, the lower-order ABs remain maximal for their order
and are denoted by broken lines. As $a\!\rightarrow\! 1/2$,
(\ref{htmax}) smoothly yields $|\psi|_{max}\!=\!1\!+\!2n$,
which are the peak heights of $n$th-order {\it rational} rogue waves (RWs) \cite{akh09}.
Thus, RWs are the natural end points of our periodic ABs.
Although RWs have the highest intensity at each 
order, their intensities are discrete, with ever-growing gaps between
successive orders. By contrast, the intensity of our hierarchy of periodic 
AB$n$, as shown in Fig. \ref{rwh}, can be continuously chosen
by changing the periodic length via
the modulation parameter $a$. 
 
Now that we have shown that this hierarchy of high-order ABs is of maximal intensity,
the next step is to find ways of producing them {\it systematically}. Currently, 
only breathers up to the second-order have been observed in optical fibers \cite{kib10,fri13,fri14}. 
While third-order breathers have been seen in random field searches \cite{dud14,toe15}, 
the analytical initial wave functions used for exciting a second-order \cite{fri13} RW optically 
or a third-order \cite{akh092} RW theoretically, were essentially obtained by trial and error. 

Recall that the AB wave function at $t\!=\!0$, (\ref{abk}), is an {\it even} function of $x$. 
Since the NLSE preserves the symmetry of the wave function, the full wave function 
must remain spatially symmetric \cite{akh86}, in the form of
\be
\psi(t,x)=A_0(t)+2\sum_{m=1}^{\infty}A_m(t)\cos(m\w x),
\ee
with complex amplitudes $A_m(t)$. As shown in Ref. \cite{chin15}, 
the instability of the fundamental mode $A_1(t)$ induces a Cascading Instability of 
all the $|m|\!\!>\!\!1$ modes, causing all to grow exponentially in locked-step with $A_1(t)$,
as $|A_m(t)|\!\sim\! |A_1(t)|^{|m|}$. Therefore, at a long 
time {\it before} the AB peak, all higher-mode amplitudes are exponentially small, 
as compared to $A_1(t)$, and the wave function must be of the form
\be
\psi_0(x)=A_0+2A_1\cos(\w x),
\la{rw1wf}
\ee
with complex amplitudes $A_0\sim 1$ and $A_1\sim 0$.
Similarly, for an $n$th-order AB, with $n$ unstable modes, the wave function at
a long time before the peak must be of the form
\be
\psi_0(x)=A_0+2\sum_{m=1}^n A_m\cos(m\w x),
\la{iwfn}
\ee
with $n$ complex coefficients $A_m$ shaping the growth of the $n$
unstable modes into a single $n$th-order AB.
Clearly, any trial and error, or grid-search method would be impractical for 
determining more than two $A_m$ coefficients.

 Here, we propose an extremely simple, yet systematic way of
 determining these coefficients. The method is to use the Darboux transformation 
to generate a numerical $n$th-order AB wave function
 at a sufficient long time before the peak and extract the $n$ coefficients $A_m$
 by {\it fitting} it with the functional form (\ref{iwfn}). (The constant $A_0$ is fixed by
 normalization.) 

\begin{figure}[hbt]
\vspace{-10 pt}
\includegraphics[width=0.95\linewidth]{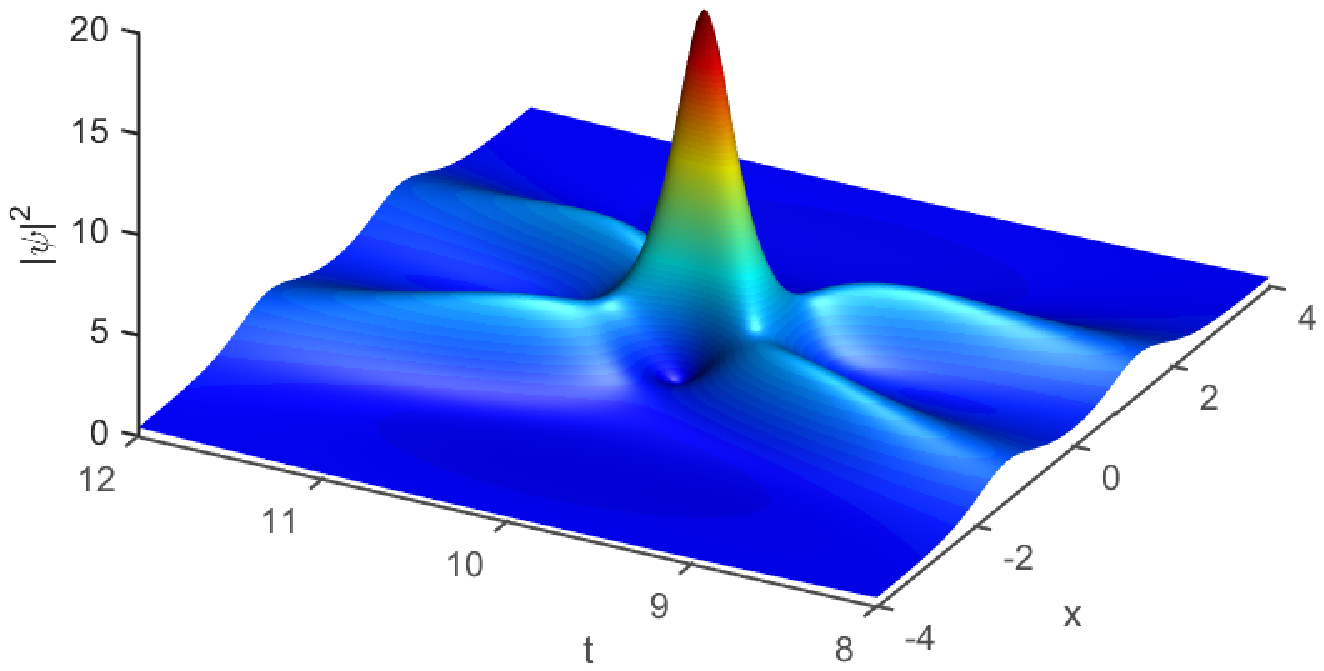}
\vspace{-10 pt}
\caption[]{\label{rwden} (color online)
 Second-order Akhmediev breather (AB) at $a=0.43$ generated from the nonlinear Schr\"odinger equation 
 using initial wave function (\ref{iwfn}). Coefficients are fitted from the Darboux transformation
 at $t\!=\!10$ before the peak;  $A_1\!=\!(0.532\!+\!1.32i)\!10^{-3}$, $A_2\!=\!(-7.56-6.54i)\!10^{-5}$, 
 $|\psi|^2_{max}$=17.48 (17.48). The value in parentheses gives the maximum intensity according 
to Eq.(\ref{htmax}).
 }
\end{figure}
\begin{figure}[hbt]
\vspace{-10 pt}
\includegraphics[width=0.95\linewidth]{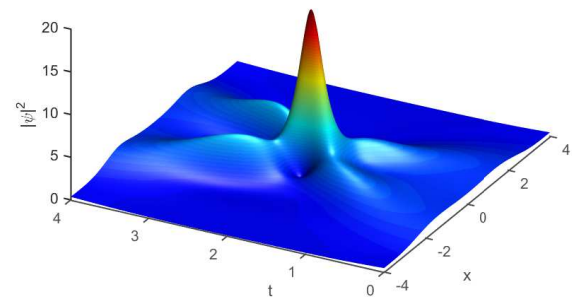}
\caption[]{\label{rwden2} (color online)
 Same as Fig. \ref{rwden}, but with coefficients fitted at $t=2$ before the peak, with only two decimal places,
 $A_1=0.18+0.28i$, $A_2=-0.11-0.03i$, $|\psi|^2_{max}$=18.61 (17.48). This second-order AB is asymmetric.
 }
\end{figure}
 
 Figures \ref{rwden} and \ref{rwden2} show the resulting second-order
 AB  at $a=0.43$ produced from the NLSE with the $n=2$ initial wave function (\ref{iwfn}).
The NLSE was solved numerically using a second-order splitting fast Fourier method
with time step $\Delta t$=0.0001 and double-checked using a fourth-order 
symplectic splitting scheme \cite{yos90}. 
We extracted the coefficients by fitting (\ref{iwfn}) to the DT wave function at $t$=$-$2 and 
at $t$=$-$10. (Therefore, when solving the NLSE numerically, the peak appears at $t=2$ and $t=10$
simulation time.) Since an overall phase is irrelevant, we subtract the phase of $A_0$ from all
coefficients, so that $A_0$ is real and we renormalize it, to obtain 
$A_0\!=\!\sqrt{1-2|A_1|^2-2|A_2|^2}$. Thus, only two
complex $A_1$ and $A_2$ are sufficient.

\begin{figure}[hbt]
\vspace{-10 pt}
\includegraphics[width=0.95\linewidth]{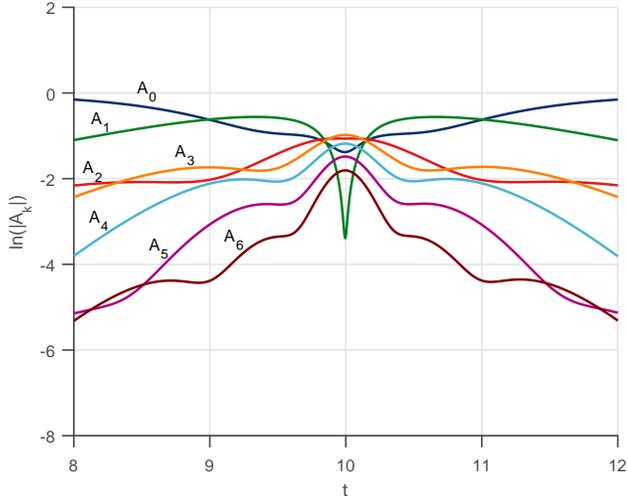}
\vspace{-10 pt}
\caption[]{\label{spect} (color online)
 The spectral ``fingerprint'' of Fig. \ref{rwden}, with coefficients fitted 
 at $t\!=\!10$ before the peak. The amplitudes are perfectly symmetric before and after the peak.
 }
\end{figure}

\begin{figure}[hbt]
\vspace{-10 pt}
\includegraphics[width=0.95\linewidth]{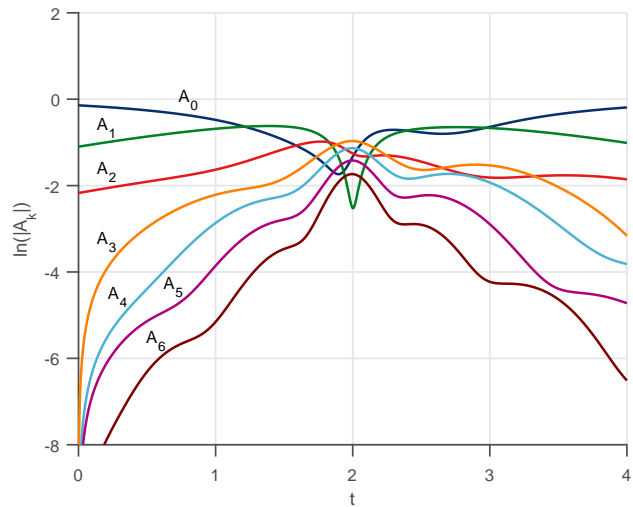}
\caption[]{\label{spect2} (color online)
The spectral ``fingerprint'' of Fig. \ref{rwden2}, with coefficients fitted at $t\!=\!2$ before the peak.
This is a poorer, asymmetric imitation of Fig. \ref{spect}, but the resulting AB still has 
comparable (actually, slightly higher) peak intensity.
 }
\end{figure}

The fitted coefficients from $t$=$-$10 generate a nearly-perfect reproduction of the AB2
generated from DT, with symmetric two-lobes before and after the peak. 
The spectral ``fingerprint'' shown in Fig. \ref{spect} is indistinguishable from the exact DT spectrum.
The fit at $t\!=\!-2$ yields larger coefficients and produces a rather distorted/asymmetric  
two-lobe structure in Fig. \ref{rwden2} and an asymmetric spectral fingerprint in Fig. \ref{spect2}. 
Yet, despite such a distortion, the latter AB2 has slightly higher peak intensity than
predicted by (\ref{htmax}).


The use of DT to analyze numerical simulations and experiments has been
done by Erkintalo {\it et al.} \cite{erk11} at the same value of $a\!=$0.43 (see their Fig. 1.) 
However, they used the $t$-shift parameter in DT to displace the two AB, so
that they only get a 1-2 peak structure, rather than a AB2. One {\it cannot}
reproduce an AB2, unless one uses the initial wave function of the form (\ref{iwfn}).

\begin{figure}[hbt]
\includegraphics[width=0.95\linewidth]{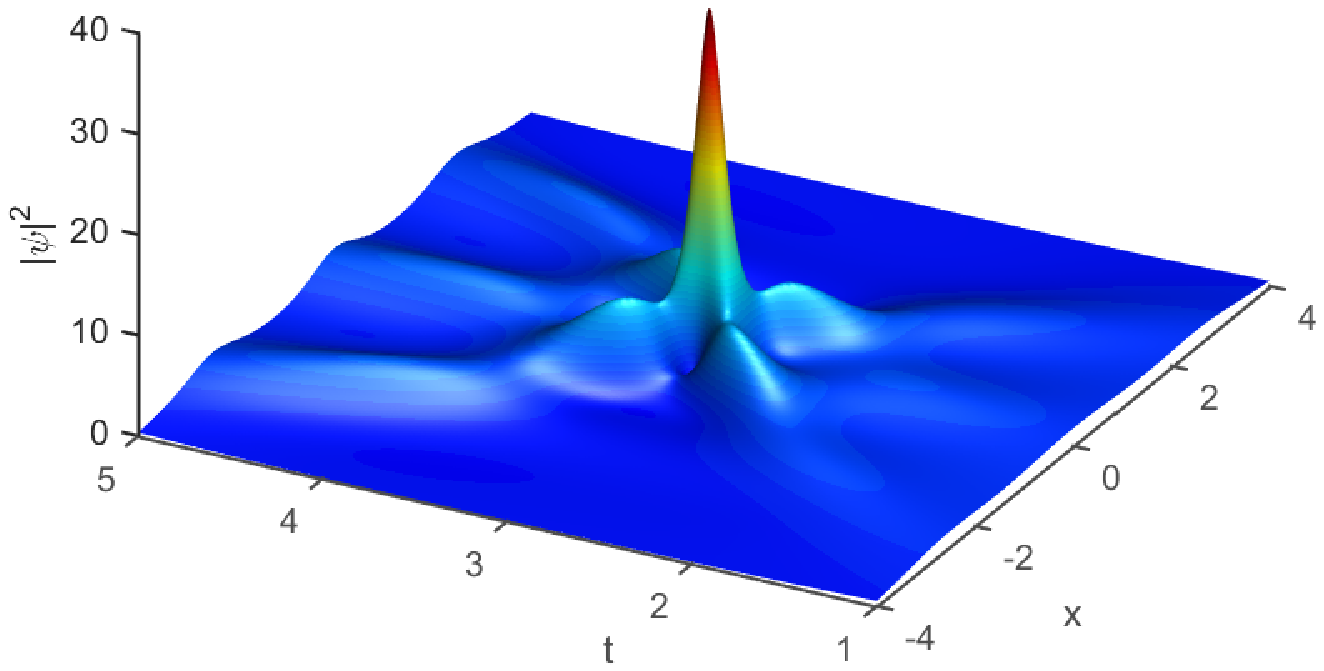}
\caption[]{\label{rw35} (color online)
 Third-order AB at $a=0.464$ from
 the initial wave function (\ref{iwfn}). Coefficients are fitted
 at $t=3$ before the peak: $A_1=0.17+0.32i$, $A_2=-0.14+0.004i$, $A_3=0.04+0.001i$,
$|\psi|^2_{max}$=35.21 (33.65). The pre-peak 3 lobes are much reduced. 
}
\end{figure}
\begin{figure}[hbt]
\includegraphics[width=0.95\linewidth]{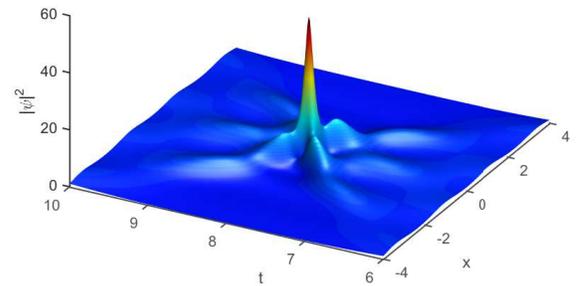}
\caption[]{\label{rw4} (color online)
  Fourth-order AB at $a=0.47395$ from
 the initial wave function (\ref{iwfn}). Coefficients are fitted
 at $t=8$ before the peak; $A_1=0.016563 + 0.067661i$, $A_2=-0.005927 - 0.005156i$, $A_3=0.002951 + 0.001122i$, $A_4=-0.008055 - 0.003309i$,
$|\psi|^2_{max}$=48.57 (49). Outer 4 lobes are at $t<6$.
}
\end{figure}
\begin{figure}[hbt]
\includegraphics[width=0.95\linewidth]{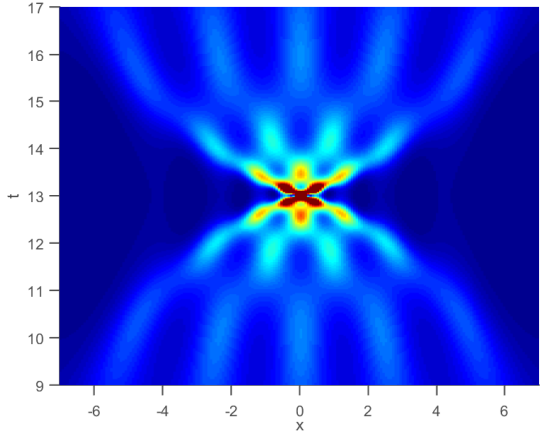}
\caption[]{\label{rw5} (color online)
 Fifth-order AB at $a\!=\!0.4850173$ from
 initial wave function (\ref{iwfn}). Coefficients are fitted
 at $t=13$ before the peak; 
$A_1=(0.64382 + 3.6195i)10^{-2}$, 
$A_2=(-1.2676 - 1.3896i)10^{-3}$,
$A_3=( 1.7862 + 1.5434i)10^{-4}$, 
$A_4=( -6.6953 - 4.4011i)10^{-5}$,
$A_5=(1.2706 + 0.6913i)10^{-4}$.
Peak intensity is
$|\psi|^2_{max}$=80 (81), but only the
base with $|\psi|^2\!<\!10$ is plotted to show the
time-symmetric 5-4-3-2-1-lobe structure.
}
\end{figure}

In Fig. \ref{rw35} we show the resulting AB3 at $a=0.464$, a value
used in the experiment of Ref. \cite{erk11}. In Figs. \ref{rw4} and \ref{rw5},
we show the resulting intensities of an AB4 and an AB5. For these two breathers,
one must fit (\ref{iwfn})  at $t$=$-$8 and $t$=$-$13 respectively, yielding rather small coefficients.

Since an $n$th-order AB is composed of $n$ 
ABs with wave numbers $\w$, $2\w$, $\cdots$ $n\w$, 
each having 1, 2, 3, $\cdots$ $n$ peaks respectively, it is equivalent to $1+2+\cdots n=n(n\!+\!1)/2$ 
single-peak ABs. This is also the observation of Ref. \cite{ank11,ked112} on rogue waves. 
This composition can be seen in the evolving intensity of all $n$th-order ABs in each region $R_n$,
not just in rational RWs \cite{ked112}. The $n$th-order AB will emerge from
the background with $n$ lobes, then ($n$-1) lobes, ($n$-2) lobes, etc., until the intensity converges 
into a narrow single peak. It then decays in a time-symmetric manner back into 2 lobes, 
3 lobes, $\cdots$ $n$ lobes, and fades back into the background. 
In Fig. \ref{rw5}, we only plot the intensity near the base of the AB, to
better show the evolving lobe structure described above.

Since the NLSE can model well the propagation of light pulses in an optical fiber, 
the above numerical generation of high order ABs strongly suggests they can also
be produced in experiments similar to those described in Refs. \cite{erk11,fri14}. The
latter's frequency-comb can basically produce all the initial wave functions given above.

\begin{acknowledgments}
This research is supported by the Qatar National Research Fund (NPRPs 5-674-1-114 and 6-021-1-005), a member of the Qatar Foundation. 
S.N.N. acknowledges support from the Serbian  MESTD Grants III45016 and OI171038. M.R.B. acknowledges support by the Al-Sraiya Holding Group.
\end{acknowledgments}

\vskip 0.2 in
\appendix*{ \centerline{{\bf Appendix: Proof of (\ref{htmax})}} }
\vskip 0.1 in

We follow the Darboux iteration in the Appendix of Ref. \cite{ked112},
with zero $x$- and $t$-shift parameters.
The wave function at $x=0$ and $t=0$ 
can be evaluated starting from their Eq.(A4), 
$$r_{1j}=2i\sin(A_{jr}),\quad s_{1j}=2\cos(B_{jr}),$$
$$
A_{jr}=\frac12\arccos(\frac{\w_j}{2})-\frac{\pi}{4},\quad B_{jr}=-\frac12\arccos(\frac{\w_j}{2})-\frac{\pi}{4},
$$
with $\w_j=2\sqrt{1-2a_j^2}$. Therefore,
\ba
s_{1j}&=&2\cos(B_{jr})= 2\cos(A_{jr}+\frac{\pi}{2})=-2\sin(A_{jr})\nn\\
&=&i r_{1j}.
\la{reqs}
\ea
Equation (\ref{reqs}) is the only result we needed to prove our formula. It follows that for all $j\ge 1$
\be
|s_{1j}|^2=|r_{1j}|^2.
\la{eqsq}
\ee
From Ref. \cite{ked112}'s Eq. (A6), 
\ba
|\psi|_{max}&=&1+\frac{2(l_1^*-l_1)s_{11}r^*_{11}}{|r_{11}|^2+|s_{11}|^2}
=1+\frac{2(l_1^*-l_1)i|r_{11}|^2}{|r_{11}|^2+|s_{11}|^2}\nn\\
&=&1+(l_1^*-l_1)i=1+(-i\sqrt{2a_1}-i\sqrt{2a_1})i\nn\\
&=&1+2\sqrt{2a_1},
\la{rw1}
\ea
where $l_n=i\sqrt{2 a_n}$. Note that we only need to know (\ref{reqs}) and (\ref{eqsq}) 
to arrive at (\ref{rw1}); we do not need to know the {\it explicit} forms of $s_{11}$ and $r_{11}$.

We now prove by induction that  (\ref{reqs}) generalizes to all $n\ge 1$, for $j\ge 1$:
\be
s_{nj}=ir_{nj}.
\la{geq}
\ee
Assumming that $s_{n-1,k}=ir_{n-1,k}$ for all $k$, specifically $k=1$, then
Ref. \cite{ked112}'s Eq. (A7) gives,
\ba
r_{nj}&=&-\sqrt{2a_{n-1}}\,s_{n-1,j+1}+i\sqrt{2a_{j+n-1}}\,r_{n-1,j+1},\nn\\
s_{nj}&=& \sqrt{2a_{n-1}}\,r_{n-1,j+1}+i\sqrt{2a_{j+n-1}}\,s_{n-1,j+1}.\nn
\ea
Now invoke $s_{n-1,k}=ir_{n-1,k}$ for $k=j+1$ then gives
\ba
ir_{nj}&=&-i\sqrt{2a_{n-1}}\,s_{n-1,j+1}-\sqrt{2a_{j+n-1}}\,r_{n-1,j+1},\nn\\
&=&\sqrt{2a_{n-1}}\,r_{n-1,j+1}+i\sqrt{2a_{j+n-1}}\,s_{n-1,j+1}\nn\\
&=&s_{nj}
\ea
From Ref. \cite{ked112}'s Eq. (A8), each $a_n$ will only contribute a factor $2\sqrt{2a_n}$ to the maximum peak 
height by applying $s_{n1}=ir_{n1}$, as done similarly in (\ref{rw1}).


\end{document}